# Forecasting with Neural Networks:
# A comparative study using the data of emergency service

Muhammad Noor-Ul-Amin

Department of Statistics, Virtual University of Pakistan

**Abstract:**

This is a case study discussing the supervised artificial neural network for the purpose of forecasting with comparison of the Box-Jenkins methodology by using the data of well known emergency service Rescue 1122. We fits a variety of neural network (NN) models and many problems were revealed while fitting the ANNs model to achieve the local minima. Moreover ANNs model is giving much better out of sample forecasts as compare to the ARIMA model. However we use diagnostic checks for the comparison of models.

## INTRODUCTION:

A lot of work has been written in the field of forecasting for the time series data. The Box Jenkin's technique has been widely used for the creation of forecasting models for linear time series. In the recent days the statisticians are efficiently using artificial neural networks as a forecasting tool. The origin of ANNs was the understanding of working of human brain and the brain networking among the neurons. Then artificial neural networking has been promoted in the field of computer sciences. An ANNs is pattern recognition or the classification of the data through a learning process. The learning process adjusts the synaptic weights that connect the neurons. The activation function is applied on the sum of product of weights and inputs in the hidden layer then the result is transferred to the output layer.

Fig 1. Functionality of NNs with one neuron using weights

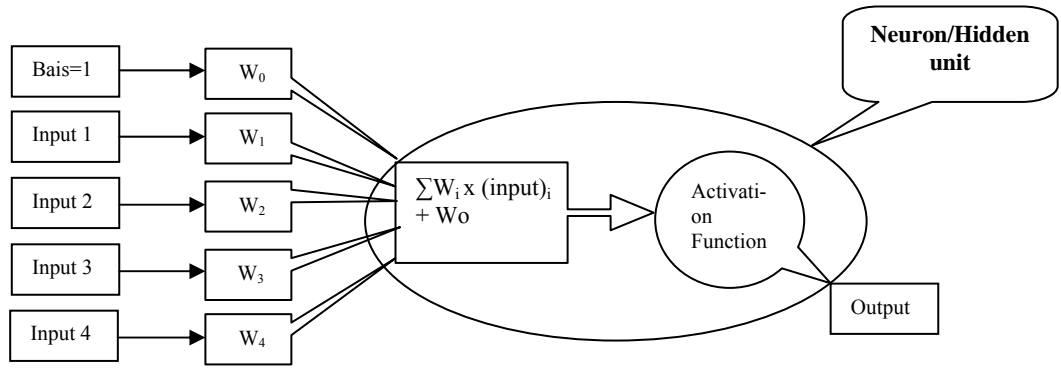

Most of the networking has been done in multilayer perception; it consists on an input layer, hidden layer(s) and an output layer. The Fig 2 describes the general architecture of the multilayer perception with the input layer consists on four neurons, hidden layer consists on 3 neurons and the output layer consists on one neuron. More than one hidden layers can be defined in multilayer perception depending on the nature of the problem. The weights are assigned to the weights to the inputs before applying the activation function in the hidden layer. The multilayer perceptron can use more than one hidden layers, then the result is transferred to the output layer.

Fig 2. Architecture of NN model using Multilayer Perception

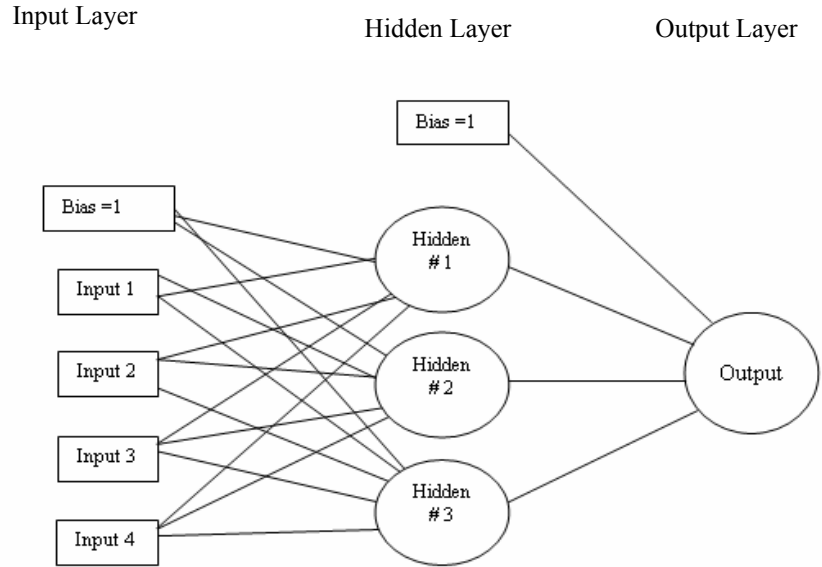

The following is the mathematical representation of the ANN modeling.

$$y_t = \phi[w_0 + \sum_{j=1}^{r} w_j f(w_{0j} + \sum_{i=1}^{k} w_{ij} \cdot y_{t-i})] + E_t$$

Where, '$y_{t-i}$' are the inputs in the model, number of hidden nodes is denoted by 'r', number of input nodes is denoted by 'k', '$w_0$' is the weight of bias term,'$w_j$' ( j = 1, 2, 3, …., q) and $w_{ij}$ ( i= 1, 2, 3, …., p, j = 1, 2, 3, …., q) are the connection weights, hidden nodes to output nodes and input nodes to hidden nodes respectively. '$f(x)$' and '$\phi(x)$' are the activation functions for the hidden layer and output layer respectively. The '$E_t$' is the error term in the model. The considerable points to obtain the successful neural network model are:

a) Nature of the data
b) Number of the hidden nodes
c) Number of the layers
    Output Layer
d) The methodology for the selection of inputs in the network
e) The activation functions for input layer and output layer
f) Training of the data
g) Optimization Algorithm

There are some guidelines are defined for (b) as the number of hidden nodes can be '2n+1' (Lippmann, 1987), 'n' (Tang and Fishwick, 1993), but the most common way for the number of hidden nodes and (c) number of hidden layers is through the experiment or trail and error (G.Zhang, 1998). The combination of hidden nodes and layers would be selected for which the error is minimum in case of test data and if the test data is not used then the network having the minimum BIC value is selected. The network which produces minimum BIC for training data is having the best number of hidden units.

About the (d), it is easy to find the inputs for the causal forecasting but in case of univeriate time series, we don't have any acceptable systematic rule to include the number of lags (G.Zhang, 1998). It has been mentioned that the inputs are same as the autoregressive terms in Box-Jenkins model (Tang and Fishwick, 1993). Tang and Fishwick concept has been refused by giving argument that AR terms define the linear relationship to the lag observations where ANN is a non-linear technique (G.Zhang, 1998). The different types of activation functions can b used. The logistic function and hyperbolic functions have been widely used in the literature. The M.Khashei (2010) suggested widely used logistic and hyperbolic activation functions for the input layer as the ANN is the non-linear mapping for the future values on the basis of past observations. The type of the activation function is depending on the situation of the neuron (M.Khashei, 2010).

Training of the data is the most important concept to understand as all the tools used within the network get into functionality under the training. Training is actually the way of preceding the records into the network. In order to train the network, we adjust the weights (increase or decrease) in such a way that difference between the network output and actual output should reduce. For measuring the change in error, the network takes the error derivatives of weights. Back propagation is a popular algorithm for calculating the derivatives. These derivatives measure the change in the error by having the change in weights and training stops when we get no change/reduction in error by changing the weights or any of the stopping rule meets. The literature recommended the three major types of the training. (1) Batch, (2) Mini Batch, (3) Online. (1) Batch uses all the records in the training data set and updates the weights when all the records have been passed once. This type of training is slow as it calculated the error as a whole but it is useful for the small data sets. (2)In mini Batch the training data is divided into groups then updates the weights by passing a group. This is faster than the Batch training. (3) Online training procedure uses one record of the training data set at a time and updates the weights by passing the each record. It is preferably used for the large training data sets. It works more quickly as compared to the batch and mini batch trainings.

The initial learning rate and momentum term is used as training options by gradient descent optimization algorithm. If the data is complex, the high momentum with low learning rate should be used and high learning rate considered for less complex data (Tang et al, 1991). G. Zhang et al. (1998) reports that these are inconsistent conclusions and suggested the best value of learning parameters are obtained by experimentation.

Fig 3. Flow chart of NN model

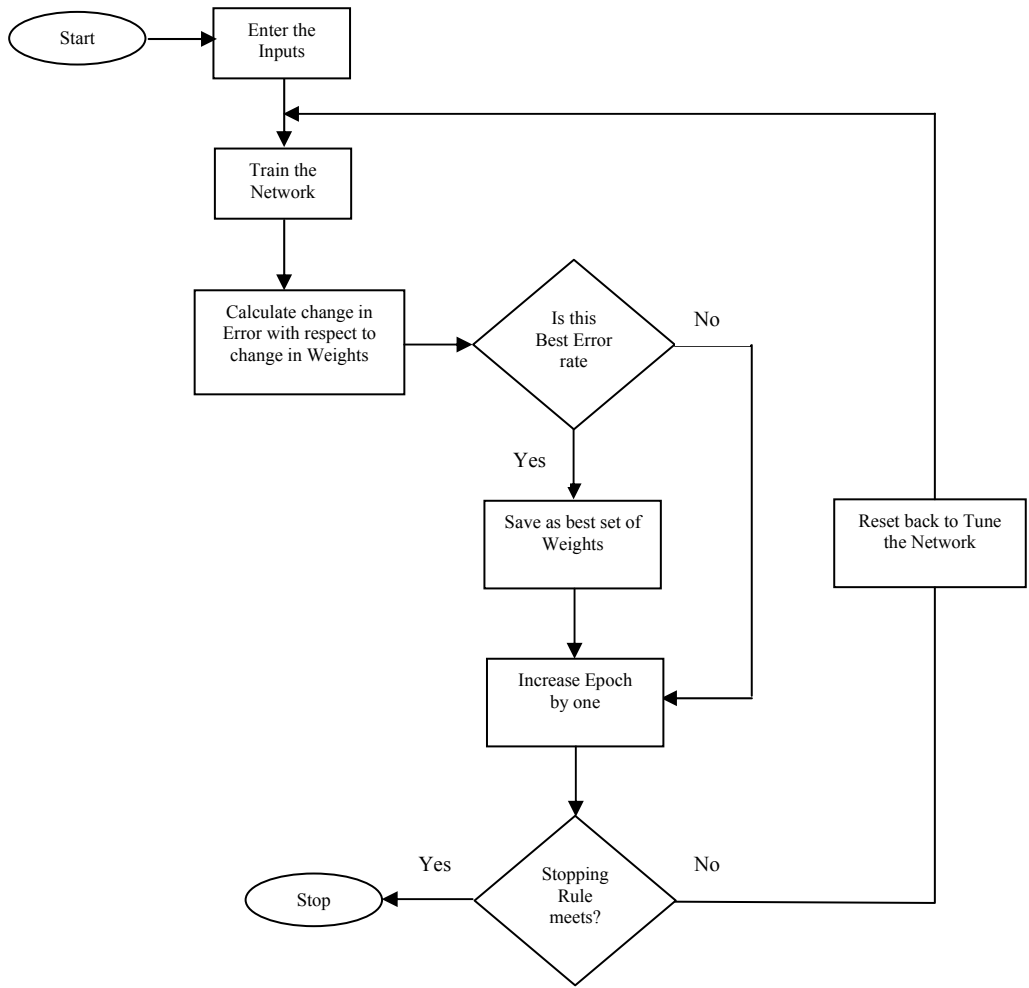

**Analysis:**

In this experiment, we take the daily call volume of emergency service Rescue 1122. The daily road accidents for the same days have also been considered as this is a cause which has an effect on the daily call volume. The 687 historical observations have been considered. The data set has been divided into two periods, first 470 in-sample values have been used for the purpose of training/analysis and rest of 217 out of sample values has been used for the accuracy testing of the obtained network/model.

Fig 4. Time Series Plot of Call Volume

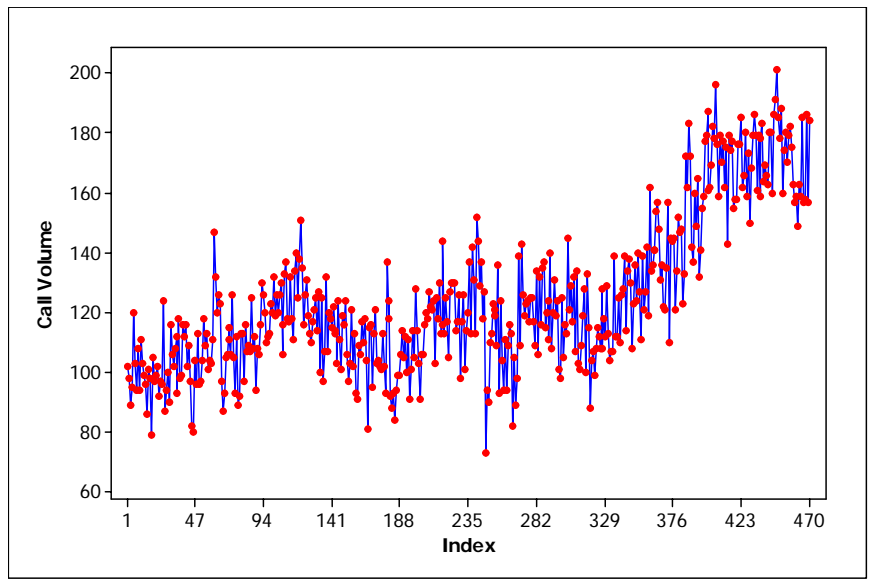

The fig. 1 is the time series plot of data. In order to make the comparison of Box-Jenkins and NNs, we fitted ARIMAX (2, 1, 1). This Box-Jenkins technique has been failed to find an appropriate model for the call volume and has been shown a low R-square 38.23% and S.S.E 76021.27.

The different order of feed forward NNs models has been tried for the data and the initial lambda value and momentum term has been defined as 0.4 and 0.9 respectively. The stopping rule for the training is to take one more step if no decrease in error is found. When analyzing the data the following major points for finding the appropriate neural network has also been addressed.

1. Number of input variables/ lag terms
2. Number of the hidden layers
3. Number of neuron in the hidden layers

The first problem was to find the appropriate input lag terms for training the network. As there have no widely acceptable rules been defined for above mentioned points but one of the idea was to include the lag terms as input by observing the ACF and PACF. Since the NNs is a non-linear technique and ACF and PACF are the functions used to find the linear relationships among the lag terms so use of autocorrelation functions has not been suggested here. We tried a variety of the NN models by including different combination of lag values and found the different local minima for the different starting points of the weights even the training algorithm and the numbers of parameters are same. It is not possible to find the global minima as there could be infinite number of repetition for finding the model. We use a number of repetitions for fitting any order of NNs with the random starting values of the weights and took the best of the resulting minima in the following table.

Table 1: The relative performance of NN models

| Order of Network | no. of parameters | Artificial Function Input/Output | R-Square | S.S.E Training/Testing |
|---|---|---|---|---|
| NN(1,1+2) | 11 | Sigmoid/ Identity | 0.834602 | 52351.64/33985.54 |
| NN(1,1-3) | 19 | Sigmoid/ Identity | 0.839164 | 50907.48/31725.14 |
| NN(1,1-4) | 22 | Sigmoid/ Identity | 0.842540 | 49839.09/31062.88 |
| NN(1,1-5) | 33 | Sigmoid/ Identity | 0.844900 | 49092.15/32382.44 |
| NN(1,1-6) | 37 | Sigmoid/ Identity | 0.844089 | 49348.9/30351.49 |
| NN(1,1-7) | 71 | Sigmoid/ Identity | 0.841576 | 50144.27/35067.98 |
| NN(1,1-8) | 78 | Sigmoid/ Identity | 0.845986 | 48748.23/33041.85 |
| NN(1,1+2+5) | 19 | Sigmoid/ Identity | 0.842440 | 49868.79/31014.09 |
| NN(1,1+5) | 11 | Sigmoid/ Identity | 0.836520 | 51744.12/33584.17 |
| NN(1,1+3) | 11 | Sigmoid/ Identity | 0.846810 | 48488.94/31759.74 |
| NN(1,2+3) | 11 | Sigmoid/ Identity | 0.834556 | 52366.22/35736.9 |
| NN(1,3+4) | 11 | Sigmoid/ Identity | 0.824637 | 55505.52/35810.35 |
| NN(1,4+5) | 11 | Sigmoid/ Identity | 0.816695 | 58019.62/40453.16 |
| NN(1,1+2) | 11 | Sigmoid/ Sigmoid | 0.814311 | 58774.17/36413.02 |
| NN(2,1-3) | 26 | Sigmoid/ Sigmoid | 0.831951 | 53190.8/33611.03 |
| NN(2,1-3) | 26 | Sigmoid/ Identity | 0.841531 | 50158.24/33335.32 |
| NN(1,1-3) | 19 | Sigmoid/ Sigmoid | 0.840067 | 50621.65/33545.72 |

The different number of lags has been used as inputs in the model. The NN (1, 1-8) is the highest order among NN models, NN (1,1-8) has 78 estimated parameter values making it complex and showing the poor choice of input lags as compared to low order NN models. Clearly NN (1, 1-8) has no significant improvement in predictive performance so it is alarming to use the higher order NN models for predictions.

When we made comparison of the NN(2,1-3) model with one layer NN model having same input variables has been shown the good fit to the training data but poor out of sample predictive performance of NN(2,1-3) so this has been suggested to chose a model with one hidden layer to avoid complications (see fig 5). Another experiment has been performed to analyze the performance of sigmoid function at the output layer. It has been shown a high value of R-Square to the training data but S.S.E for the test data is exposing the poor out of sample forecast performance. The use of sigmoid function at the output layer has been shown the instable forecasts so it is suggested to use the identity function at the output layer. The NN (1, 1-3) is low order NN model and it's S.S.E of Training/testing data telling the story of good fit to the training data and it's good out of sample predictive performance but NN(1, 1+3) has been shown more satisfactory results. Since NN (1, 1+3) is a low order NN model with a low amount of S.S.E for testing data should be selected as the best model among all the fitted models.

Fig 5. Comparison between original values and estimated values

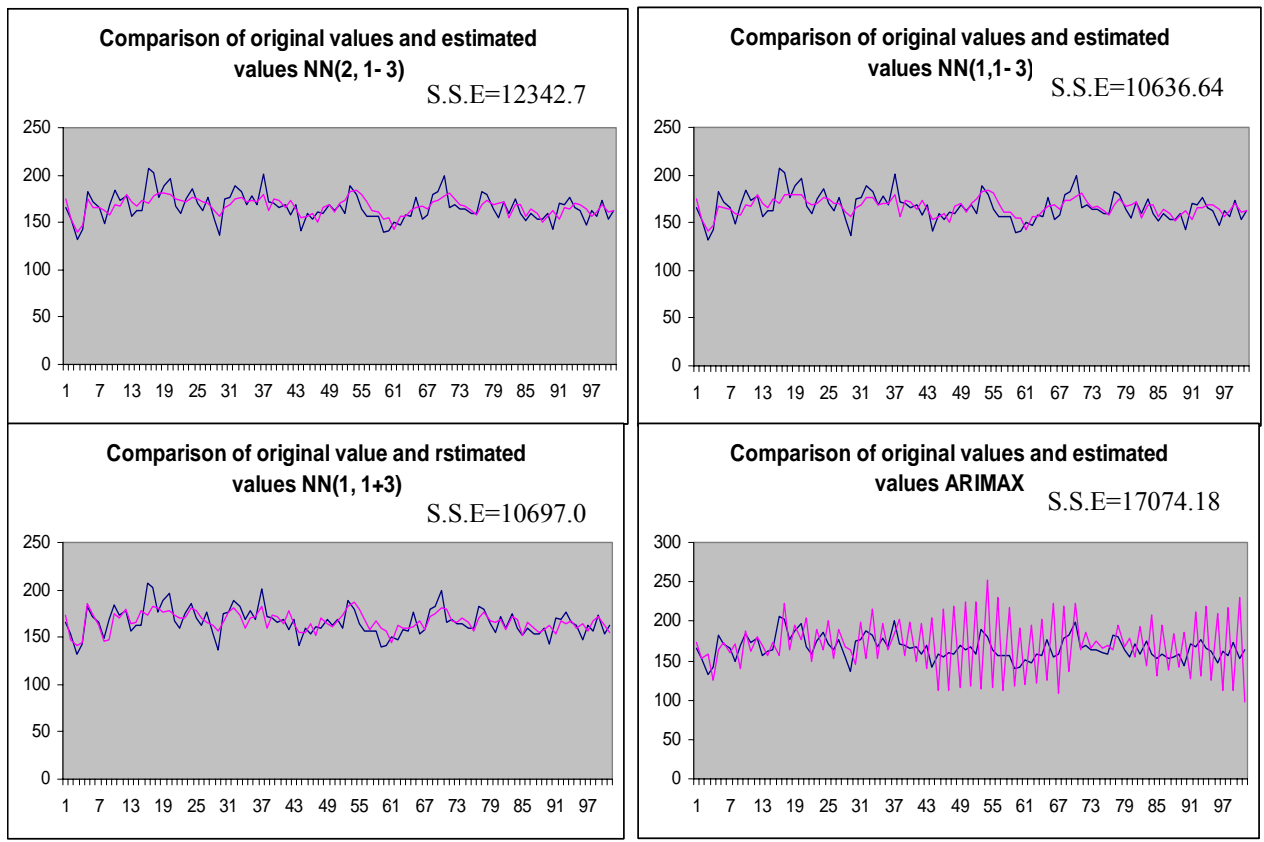

In the given fig 5, we made the comparison of forecasts for the next 100 observations. There figures for NN (2, 1-3) and NN (1, 1-3) have been shown almost the same behavior but the lower value of S.S.E for NN (1, 1-3) has been indicated batter efficiency with one layer NN model. Clearly ARIMAX prediction performance is poor than NN (1, 1+3) model. The NN (1, 1+3) has been selected as the most appropriate model in the fitted NN models.

**Conclusion:**

We suggest the following points in the conclusion

- An experience researcher who knows the art of forecasting should apply the NN models as a great care is required for selection of architecture of the NN models such as number of hidden layers, number of input variables, number of neuron, selection of activation functions, starting values of the weights, type of training, optimization algorithm, choice of learning rate and momentum term and stopping rule. For finding the appropriate model the researcher needs a rush of refitting the NN model.
- There is a plenty of scope for working on NN models and the definite rules should be defined for the architecture of NN modeling.
- In this case study of forecasting the call volume, the low order NN models are much suitable than higher order NN models and the inclusion of more lag terms has not been the reason for the improvement in forecasts.
- The inclusion of more than one hidden layers has not been the cause for improvement into the prediction performance and the use of sigmoid function on the output layer giving poor forecast as compare to the identity function.
- The Box Jenkins technique has been failed to find the suitable forecasting model, on the other hand the NN models has been given comparatively satisfactory results.

**Reference:**

1. Ahmad, Nesreen K. , Atiya, Amir F. , Gayar, Neamat Eland El-Shishiny, Hisham(2010). An Empirical Comparison of Machine learning Models for Time Series Forecasting, Econometric Reviews, 29: 5, 594-621.

2. Balkin, Sandy D. , Ord J.keith (2000). Automatic neural network modeling for univariate time series, International journal of forecasting, Vol 16, 509-515.

3. Brath A., Montanari A., and Toth E (2002). Neural networks and non-parametric methods for improving real time flood forecasting through conceptual hydrological models, Hydrology and Earth System Sciences, 6(4), 627- 640.

4. Bruce H. Andrews, Shawn M. Cunningham, L. L. Bean Improves Call-Center, Interfaces, Vol. 25, No. 6, November-December 1995, pp. 1-13.

5. Channouf, Nabil (2006). The application of forecasting techniques to modeling emergency medical system calls in Calgary, Alberta, Health Care Management Science, Vol. 10, No. 1. (February 2007), pp. 25-45.

6. Faraway, Julian, Chatfield, Chris (1998). Time series forecasting with neural networks: a comparative study using the airline data, Appl. Statist. Vol 47, Part 2, 231-250.

7. Gooijer, Jan G. De, Hyndman, Rob J. , 25 years of time series forecasting, International journal of Forecasting 22 (2006) 443-473.

8. Gunther, Frauke, Fritsch, Stefan (2010). Neuralnet: Training of Neural Networks, the R Journal Vol. 2/1, ISSN 2073-4859.

9. Haider, Adnan, Hanif, Muhammad Nadeem (2009). Inflation Forecasting In Pakistan using Artificial Neural networks, Pakistan Economic and Social Review.Vol 47, No. 1, 123-138.

10. Jordan Michael I. (1996). Neural Networks, ACM Computing Surveys, Vol 28, 73-75.

11. Khashei, Mehdi., Bijari, Mehdi (2010). An artificial neural network (p,d,q) model for time series forecasting, Expert Systems with Applications 37 479-489.

12. Lai, Kin keung, Yu, Lean, and Wang, Shouyang (2009). Neural-Network-based Metamodeling for Financial Time Series Forecasting, Applied Soft Computing, Vol 9, issue 2, 563-574.

13. Lippmann, R.P (1988). An introduction to computing with neural nets, ACM SIGARCH Computer Architecture News, Vol 16, 7-25.
14. P. Holcomb, John and Radke Sharpe, Norean (2007). Forecasting Police Calls during Peak Times for the City of Cleveland: CS-BIGS 1(1): 47-53.
15. Reynolds Penny (2005), Forecasting Fundamentals: The Art and Science of Predicting Call Center Workload, The Call Center School, Lebanon.
16. Tang Z., Fishwick, P.A(1993). Feedforward Neural nets as models for time series forecasting. ORSA Journal on Computing, Vol 5(4), 374-385.
17. Wallace Martin P.(2008). Neural networks and their application to finance, Business intelligence journal, Vol 1, 67-76.
18. Zhang, G., Patuwo, B. E., and Hu, M.Y. (1998). Forecasting with artificial neural networks: the state of the art, International journal of forecasting, Vol 14, 36-62.